\documentclass{emulateapj}
\usepackage{amssymb}
\usepackage{amsmath}
\usepackage{natbib}
\usepackage{gensymb}
\usepackage{txfonts}
\usepackage[colorlinks=true,citecolor=blue,linkcolor=blue]{hyperref}
\usepackage{microtype}
\usepackage{color}
\usepackage{graphicx}
\usepackage{epstopdf}
\usepackage{multirow}
\usepackage{tabularx}
\usepackage{lineno}
\linenumbers

\def\apjl{ApJL}
\def\apjs{ApJS}

\def\aap{A\&A}

\usepackage{soulutf8} 
\usepackage[usenames,dvipsnames]{xcolor}

\usepackage{changes}

\shorttitle{Constraining some parameters of the pulsar wind model via statistical modeling}
\shortauthors{Coelho et al.}

\begin{document}

\title{Observational Constraints on the Pulsar Wind Model: The Cases of Crab and Vela}

\author{Jaziel G. Coelho\altaffilmark{1,2}, José C. N. de Araujo\altaffilmark{2}, Samantha M. Ladislau\altaffilmark{2} and Rafael C. Nunes\altaffilmark{2}}

\altaffiltext{1}{Departamento de F\'isica, Universidade Tecnol\'ogica Federal do Paran\'a, 85884-000 Medianeira, PR, Brazil}

\altaffiltext{2}{Divis\~ao de Astrof\'isica, Instituto Nacional de Pesquisas Espaciais, Avenida dos Astronautas 1758, 12227-010, S\~ao Jos\'e dos Campos, SP, Brazil}

\begin{abstract}
As is well known, pulsars are extremely stable rotators.
However, although slowly, they spin down thanks to brake mechanisms, which are in fact still subject of intense investigation in the literature. Since pulsars are usually modelled as highly magnetized neutron stars that emit beams of electromagnetic radiation out of their magnetic poles, it is reasonable to consider that the spindown has to do with a magnetic brake. Although an interesting and simple idea, a pure magnetic brake is not able to adequately account for the spindown rate. Thus, many alternative spindown mechanisms appear in the literature, among them the pulsar wind model, where a wind of particles coming from the pulsar itself can carry part of its rotational kinetic energy. Such a spindown mechanism depends critically on three parameters, namely, the dipole magnetic field $B$, the angle between the magnetic and rotation axes $(\phi)$, and the density of primary particles $(\zeta)$ of the pulsar's magnetosphere. Differently from a series of articles in this subject, we consider for the first time in the literature a statistical modelling which includes a combination of a  dipole magnetic and wind brakes. As a result, we are able to constrain the above referred parameters in particular for Crab and Vela pulsars.
\end{abstract}


\altaffiltext{}{jazielcoelho@utfpr.edu.br \\}

\maketitle

\section{Introduction}
\label{intt}
As is well known, Pulsars, which is usually associated with rotating neutron stars (NSs), have extremely stable rotating periods. In particular the so-called rotation-powered pulsars (RPPs) emit radiation by means of their rotational kinetic energies, as a result their periods increase, i.e., they spindown~\citep{1969Natur.223..813O,1969Natur.221..454G}. The electromagnetic energy emitted by a pulsar, in this case, come from its rotational kinetic energy~\citep[see e.g.,][]{1975ctf..book.....L,2001thas.book.....P}.

It is very likely that pulsars, due to their dynamic nature, should always present important temporal changes in some astrophysical quantities. In particular, increases in the rotational periods, for example, are usually quite small. The Crab Pulsar (PSR B0531+21), for example, which has a period of $\sim 33$ ms, has a period increase rate of $\simeq 4.2 \times 10^{-13}$ s/s. While the Vela Pulsar (PSR J0835-4510 or PSR B0833-45) has a spin period of $\sim 89$ ms and a spindown rate of $\simeq 1.25 \times 10^{-13}$ s/s~\footnote{For information about pulsars, we refer the reader to the ATNF catalog available at: \url{https://www.atnf.csiro.au/research/pulsar/psrcat/}}.

A long standing issue is to understand how exactly the pulsars spindown. The magnetic dipole radiation model is a simple and interesting proposal to explain the spindown. However, such a model predict that the brake index, a dimensionless quantity that relates the period and its first and second time derivatives, is exactly equal to three, which is not observationally corroborated. In addition to that, the estimation of the dipole magnetic field is subject to several uncertainties. For instance, several analysis suggests that should not be disregard the possibility of multipolar magnetic field in highly magnetized stars~\citep[see, e.g,][]{2020ApJ...889..165D}. Indeed, NICER’s X-ray data from PSR J0030+0451 has recently led to the first map of the hot spots on the surface of a star~\citep[see][]{2019ApJ...887L..21R,2019ApJ...887L..26B}. The hot spots are far from antipodal, meaning that the magnetic field structure of a pulsar is much more complex.

The fact that no pulsar has a braking index equal to three implies the need to consider more elaborate spindown models. One such a model is the so-called pulsar wind model~\citep[see][]{2001ApJ...561L..85X,2015MNRAS.450.1990K,2017ApJ...837..117T}, which we consider in the present work. We shall see later in this paper that different values of braking index is naturally the case whenever pulsar wind mechanism also features in the energy loss budget of pulsars, along with the classic magnetic dipole radiation. In addition, observations of intermittent pulsars showed explicitly the substantial role of particle wind in pulsar spindown~\citep[see][]{2006Sci...312..549K}. Magnetohydrodynamics simulations also found similar expressions to the wind braking model~\citep[see e.g.,][]{2006ApJ...648L..51S}. In the next section, we briefly review such a model.

It is worth mentioning that there are several scenarios that challenge the classic magnetic dipole model, like the one involving the accretion of fall-back material via a circumstellar disk~\citep{2016MNRAS.455L..87C}, and modified canonical models to explain the observed braking index ranges~\citep[see][among others, and references therein for further models]{1997ApJ...488..409A,2012ApJ...755...54M,2016ApJ...823...34E,2016JCAP...07..023D,2016ApJ...831...35D,2016EPJC...76..481D,2017EPJC...77..350D}. Another interesting model for the brake is the quantum vacuum friction. We refer the reader to \cite{2016ApJ...823...97C} for details. Therefore, energy loss mechanisms for pulsars are still under continuous debate.

As already mentioned we consider here the pulsar wind model, but following a different approach than that usually adopted in the literature. By means of a statistical model, we analyze in particular three relevant parameters of the wind model, namely, $B$, $\phi$, $\zeta$, the dipole magnetic field, the initial angle between the rotation and magnetic axes, and the parameter related to the density of primary particles of the magnetosphere, respectively. 

The present paper is organized as follows. In Section \ref{PWM}, we revise the pulsar wind model, in Section \ref{SM}, we present the statistical model to analyze the parameters $B$, $\phi$ and $\zeta$ for the Crab and Vela pulsars. The results and discussions are presented in Section \ref{results}. The main conclusions are summarized in Section \ref{summary}.

\section{Pulsar Wind model}
\label{PWM}
In this section we briefly review the pulsar wind model as originally put forward by~\citet{2001ApJ...561L..85X} in order to elucidate the physical ideas involved. 

Let us consider the pulsar as an oblique rotator that has two components of magnetic dipole: one parallel and other perpendicular to the axis of rotation of the pulsar. The perpendicular component is responsible for the energy loss by the magnetic dipole radiation~\citep[see e.g.,][]{1975ctf..book.....L,2001thas.book.....P}, whereas the parallel component is related to the acceleration of particles~\citep[see][]{2014ApJ...788...16L}. Then, the phenomenon of pulsar wind is basically an energy loss mechanism due to the classic magnetic dipole radiation and particle
acceleration~\cite[see][]{2015MNRAS.450.1990K,2017ApJ...837..117T}.

The energy loss due to particle wind depends on the so-called acceleration potential drop, $\Delta v$, given by~\citep{2001ApJ...561L..85X}
\begin{equation}
    \dot{E}_{\rm wind}=2\pi r^2_p c \rho_e \Delta v,
\end{equation}

\noindent where $r_p= R(R\Omega/c)^{1/2}$ is polar gap radius, $c$ is the speed of light, $\rho_e$ is the primary particle density, and  $\Delta v$ is the corresponding acceleration potential in the acceleration gap. The density of primary particles is related to the  Goldreich–Julian charge density by $\rho_e=\zeta\rho_{GJ}$~\citep{1969ApJ...157..869G}, being $\zeta$ a coefficient which can be constrained by observations. It is important to note that $\zeta$ is related to the primary particles in the acceleration gap but not to the total outflow particles. 

Notice that the presence of the acceleration potential can accelerate primary particles. Secondary particles are generated subsequently. Meanwhile, the density of secondary particles can be much higher than the Goldreich–Julian density. In the wind braking model, all the particles injected into the magnetosphere from the acceleration region are defined as primary particles.

If we assume that the maximum potential for a rotating dipole is given by $\Delta V=\mu\Omega^2/c^2$, it can be shown that the rotational energy loss rate reads
\begin{equation}
     \dot{E}_{\rm wind}=\frac{2\mu^2\Omega^4}{3c^3}3\zeta\frac{\Delta v}{\Delta V}\cos^2\phi,
\end{equation}

\noindent where $\mu = 1/2 BR^3$ is the magnetic dipole moment ($B$ is the magnetic field strength at the magnetic pole of the star and $R$ is the neutron star radius), $\Omega$ is the rotational frequency, and $\phi$ the inclination angle between rotation and magnetic axes.

On the other hand, as it is well known, pulsars also lose energy via the classic magnetic dipole radiation~\citep[][]{2001thas.book.....P,1975ctf..book.....L}.
The magnetic dipole radiation and the outflow of particle wind may contribute independently. Then, the total rotational energy loss rate is given by \citep{2015MNRAS.450.1990K,2017ApJ...837..117T}
\begin{equation}
    \dot{E}=\frac{2\mu^2\Omega^4}{3c^3}\left(\sin^2\phi+3\zeta\frac{\Delta v}{\Delta V}\cos^2\phi\right)=\frac{2\mu^2\Omega^4}{3c^3}\chi.
    \label{RotEnerLoss}
\end{equation}

We note that if the acceleration potential $\Delta v=0$, there are no particles accelerated in the gap, the pulsar is just braking down by
the magnetic dipole radiation. Here $\chi$ is a dimensionless
function that can be viewed as the dimensionless spin-down
torque. The expressions of $\chi$ for different acceleration models had been very well studied by \citet{2015MNRAS.450.1990K}
[see Table 2 therein for various acceleration models]. In fact, the $\chi$ parameter depends on the particle acceleration model adopted. Here, we will use the vacuum gap (VG) model with curvature radiation (CR)~\citep[see][]{1975ApJ...196...51R}.

We shall surmise in this work that the total energy of the star is provided by its rotational counterpart, $E_{\rm rot}=I\Omega^2/2$, and its change is attributed to both $\dot{E}_{\rm wind}$ and the magnetic dipole radiation. Thus, from Eq~\ref{RotEnerLoss}, the evolution of the rotational frequency of a star is given by

\begin{equation}
     \dot{\Omega} =  -\frac{B^2R^6 \Omega^3}{6 Ic ^3}~\chi^{CR}_{VG},
     \label{rotfreq}
     \end{equation}

\noindent with

\begin{equation}\label{morte}
\chi^{\rm CR}_{\rm VG}
= \sin^{2}\phi + 
\left\{
\begin{array}{l}
4.96\times 10^{2} \zeta \left(1-\frac{\Omega_{\rm death}}{\Omega}\right) B_{12}^{-8/7}\Omega^{-15/7}  \\
\mbox{\hspace{90pt}if $\Omega > \Omega_{\rm death}$}  \\
0 
\mbox{\quad if $\Omega < \Omega_{\rm death}$,} \, \end{array} \right.
\end{equation}

\noindent where the term in parentheses account for the pulsar death,   
and $\rm B_{12}$ is the surface magnetic field in units of 10$^{12}$ G.
Notice that in the above equation, the term $\cos^2\phi$, which appears in Eq \ref{PWM}, is now omitted. In \cite{2017ApJ...837..117T}, the authors argue that $\cos^2\phi$ may not appear, in accordance with magnetospheric simulations performed by \cite{Li_2012}.

Consequently, the effect of pulsar death can be incorporated in the rotational energy loss rate and must be considered in modelling the long-term rotational evolution of the pulsar. Note that when a pulsar is dead ($\Omega < \Omega_{\rm death}$), it is braked only by magnetic dipole radiation, i.e., $\chi^{CR}_{VG}=\sin^2\phi$. Then, following the same procedure of \citep{2006ApJ...643.1139C,2015MNRAS.450.1990K}, the death period ($P_{\rm death}=2\pi/\Omega_{\rm death}$) is defined as
\begin{equation}
    P_{\rm death}=2.8\left(\frac{B}{10^{12} G}\right)^{1/2}\left(\frac{V_{gap}}{10^{12}V}\right)^{-1/2} s.
\end{equation}

The inclination angle $\phi$ is allowed to evolve over time, and following \cite{2017ApJ...837..117T}, the evolution of $\phi$ reads

\begin{equation}
    \dot{\phi} = - \frac{B^2R^6\Omega^2}{6 I c^3}\sin{\phi}\cos{\phi}
\label{dphi}
\end{equation}

As already mentioned, the energy carried away by the dipole  radiation and the relativistic particles originates from the rotational kinetic energy, the loss rate of which is $I\Omega\dot{\Omega}$.

Recall that the braking index is defined by,
\begin{equation}
n = \frac {\Omega \ddot \Omega} {{\dot \Omega}^2}.
\label{BI}
\end{equation}

It is interesting to note that the braking index implicitly depends on the magnetic field $B$, the inclination angle $\phi$, and particle density $\zeta$.

\section{Statistical model for Crab and Vela}
\label{SM}

In \cite{2015MNRAS.450.1990K} and, in particular and mainly in \cite{2017ApJ...837..117T}, one sees that the key parameters to appropriately model the pulsar spindown when considering a combination of magnetic dipole and particle wind brakes are $B$, $\phi$ and $\zeta$.

In a modelling for the Crab pulsar, \cite{2015MNRAS.450.1990K} assume that $B=8.1\times10^{12}$~G, $\phi=55\degree$ and $\zeta=10^3$. Later, \cite{2017ApJ...837..117T} adopt $B\sim10^{12}$~G, $\phi=60\degree$ and $\zeta=10^2$. The authors argue that the primary particle density, $\rho_e$, of young pulsars is at least 80 times the $\rho_{GJ}$ in the vacuum gap model. In fact, the particle density in the accelerating region could be $\sim10^3$ to $10^4$ times the  Goldreich–Julian charge density~\citep[see also][]{2007AdSpR..40.1491Y}. A much larger particle density  than  the  Goldreich–Julian  density  in  the  pulsar magnetosphere is also found in other models and observations~\citep[and references therein]{2015MNRAS.450.1990K,2017ApJ...837..117T}. 

\begin{table*}
\caption{Period ($P$), its first derivative ($\dot P$), surface magnetic field ($B$), braking index ($n$) and spindown (SD) age for the Vela and Crab pulsars.}
\begin{ruledtabular}
{\begin{tabular}{@{}cccccccccc@{}} 
Pulsar & $P$~(s) &$\dot{P}~(10^{-13}$~s/s) &B ($10^{12}$G)$^{*}$ & n & age (kyr)$^{**}$ & Ref.   \\ \hline
PSR B0833-45 (Vela) &0.089&  1.25& 6.8 & $1.4\pm0.2$ & 11.3 &~\cite{1996Natur.381..497L,2017MNRAS.466..147E} \\
PSR B0531+21 (Crab) &0.033 &4.21 & 7.5 & $2.51\pm0.01$ & 0.967  &~\cite{1993MNRAS.265.1003L,2015MNRAS.446..857L}  \\
\end{tabular} \label{ta1}}
\end{ruledtabular} 
\\
{$^{*}$ $ B = 6.4\times 10^{19} \sqrt{P\dot P} \; {\rm G}$ - for canonical parameters of $M$, $R$ and $I$.}\\
{$^{**}$ For the Vela pulsar we use the spindown age = $ P/2\dot{P}$. However, we have adopted the true age for Crab pulsar, which is known to be just 967 yr because the Crab supernova was observed in 1054 AD.}

\end{table*}

Differently from these and other previous studies, we consider for the first time in the literature a statistical modelling which includes a combination of a dipole magnetic and wind brakes. We argue that a robust way to adequately obtain and constrain $\phi$, $\zeta$ and $B$ is by mean of statistical analysis.

According to the inferred observational range of inclination angles and characteristic magnetic fields, we are able to constraint the range of values of $\phi$ and $\zeta$ for a particular pulsar. As a first application of our modelling we consider the widely known Crab and Vela pulsars.

Here, we use the Markov Chain Monte Carlo (MCMC) method to analyze the parameters $\theta_i = {\phi, \zeta, B}$, building the posterior probability distribution function 

\begin{equation}
\label{L}
 p(D|\theta) \propto \exp \Big( - \frac{1}{2} \chi^2\Big) \, ,
\end{equation}
where 

\begin{equation}
\label{chi2_m}
\chi^2 = \Big(\frac{n - n_{\rm th}}{\sigma_{n}} \Big)^2 \,,
\end{equation}
where $n$, $n_{\rm th}$ and $\sigma_n$ are the observed braking index (median value), theoretical braking index and the uncertainties of the observed braking index~(see Table~\ref{ta1}), respectively.

The goal of any MCMC approach is to draw $M$ samples $\theta_i$ from the general posterior probability density
\begin{equation}
\label{psd}
p(\theta_i, \alpha|D) = \frac{1}{Z} p(\theta,\alpha) p(D|\theta,\alpha)  \, ,
\end{equation}
where $p(\theta,\alpha)$ and $p(D|\theta,\alpha)$ are the prior distribution and the likelihood function, respectively. Here, the quantities $D$ and $\alpha$ are the set of observations and possible nuisance parameters. The amount $Z$ is a normalization term.
In order to constrain the baseline $\theta_i$, let us assume estimates of the braking index parameters for the pulsars as follows: n = $2.51\pm0.01$ for Crab and n = $1.4\pm0.2$ for Vela (see Table \ref{ta1}). 

We perform the statistical analysis based on the \textit{emcee} algorithm \citep[see][]{2013PASP..125..306F}, assuming the theoretical model described in Sec. \ref{PWM} and the following priors on the parameters baseline: first, we analyze both Vela and Crab with a uniform prior on the inclination angle to be $\phi \in [45^{\circ}, 70^{\circ}]$, which are consistent with observational constraints~\citep{2013Sci...342..598L}. As a second case, we consider a uniform prior $\phi \in [70^{\circ}, 90^{\circ}]$.  In fact, the shape of the beam of the Crab pulsar has been investigated over the past few years, resulting in a range of estimates of $\phi \in [45^{\circ}, 70^{\circ}]$~\citep[see e.g.,][]{2003ApJ...598.1201D,2008ApJ...680.1378H,2009ApJ...695.1289W,2012ApJ...748...84D}. 
Unfortunately, at present it is impossible to accurately determine the inclination angle of individual pulsars. Therefore, these issues are still under continuous debate~\citep[see e.g.,][]{2018MNRAS.481.4169L,2020MNRAS.494.3899N}.

From the profile modeling, we can already get some information about the inclination angle. In fact, the braking index is not the only observational input, since preliminary information on $\phi$ is already known. Thus, we use this information as uniform prior in our analysis. We are fitting the theoretical model under an observational information quantified in terms of $n$, which represents in practical terms just one data point, with already known information on $\phi$. Thus, we will maintain a conservative statistical limit in our results, and we will quantify our all analysis at 38\% ($\sim$0.5$\sigma$) and 68\% ($\sim$1$\sigma$) CL. In what follows, let us present a summary of our main results.
\begin{figure}[!]
\centering
\includegraphics[width=8cm, height= 8cm]{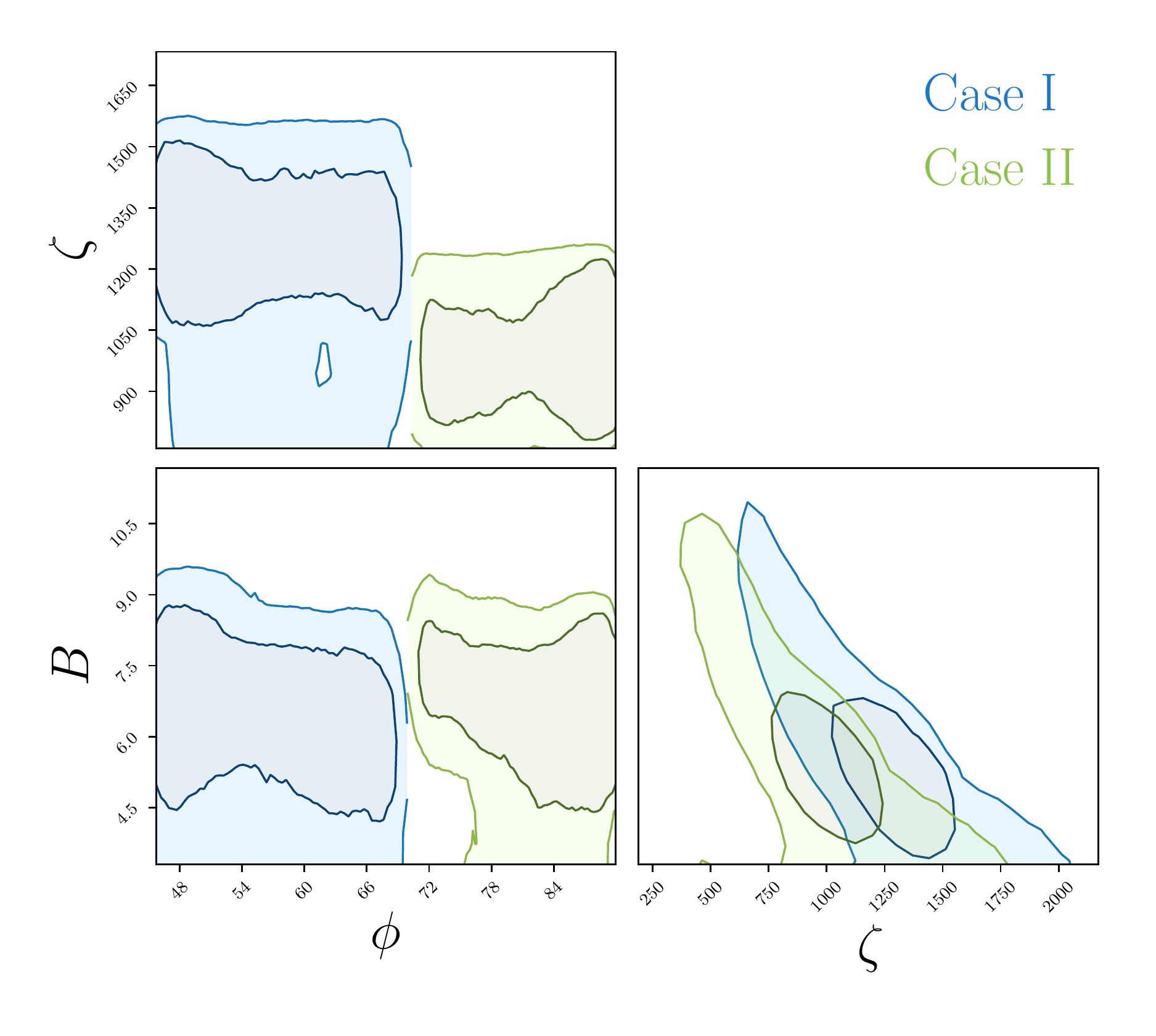} 
\includegraphics[width=8cm, height= 8cm]{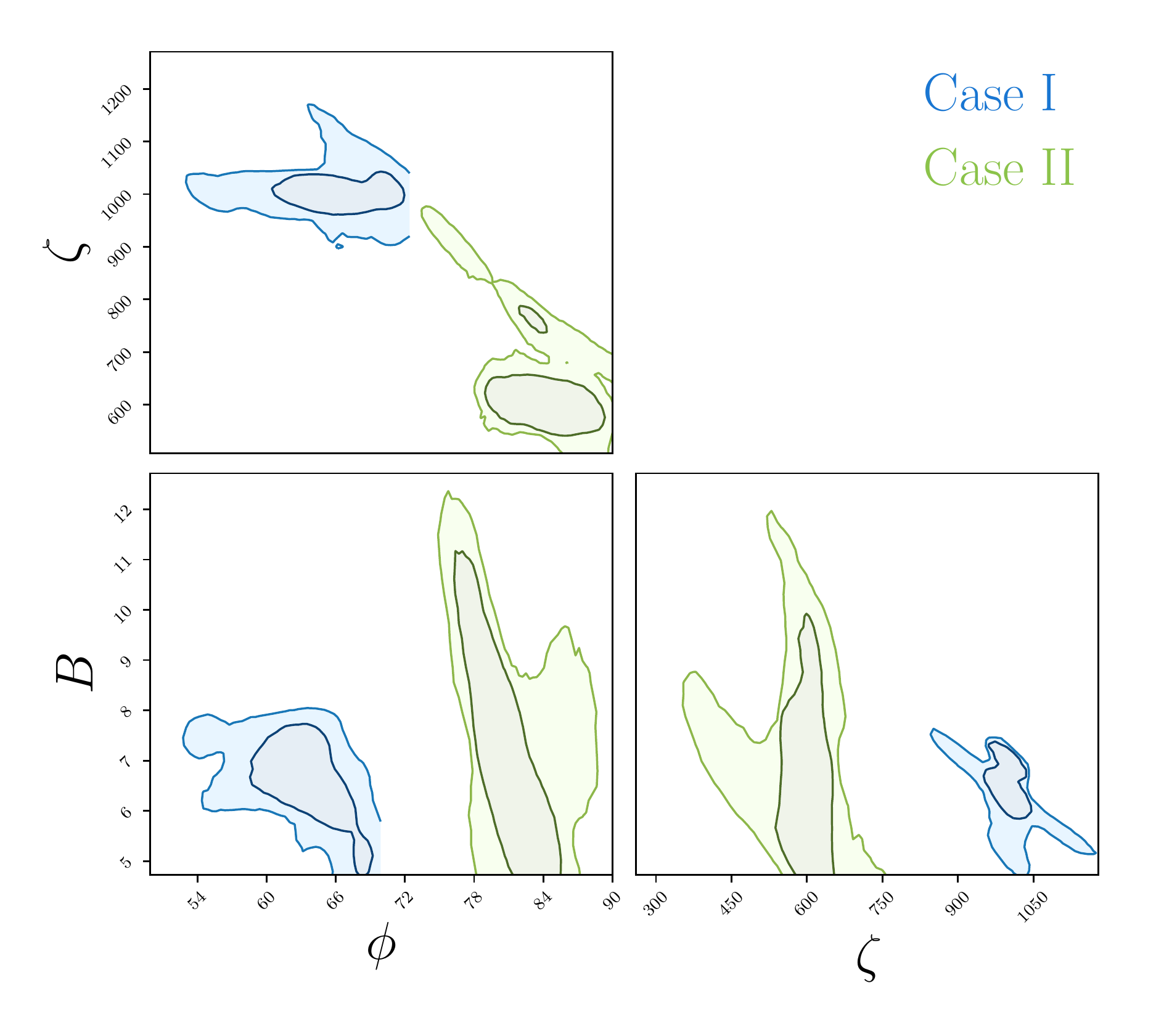} 
\caption{The parametric space at 38\% CL and 68\% CL, under the prior consideration $\phi \in [45^{\circ}, 70^{\circ}]$ (Case I) and $\phi \in [70^{\circ}, 90^{\circ}]$ (Case II). Upper panel: Vela Pulsar. Lower panel: Crab Pulsar. The parameter $B$ is in units of $10^{12}$ G.}
\label{PS_prior_70_90}       
\end{figure}

\section{Results and Discussions}
\label{results}

In the following we explore the parameter space $\phi$, $\zeta$ and $B$ with our MCMC approach, in order to constrain the probability distribution of these parameters that characterize the pulsar wind model. Then, we relaxed the value of $B$ using a uniform prior with $B \in [1, 100]$ in units of $10^{12}$ G. As a case study, in Fig. \ref{PS_prior_70_90} we show the parametric space on the plan $\phi$-$\zeta$ at 38\% and 68\% confidence level (CL), assuming $\phi \in [45^{\circ}, 70^{\circ}]$ (Case I) and $\phi \in [70^{\circ}, 90^{\circ}]$ (Case II).

The age of a pulsar is a useful parameter, but it is difficult to get the age from observations. Here, we have used the values showed in Table~\ref{ta1}. For the Vela pulsar we adopted the spindown age. This age is in good agreement with independent age estimators (e.g., proper motion and SNR age). It is worth mentioning that the different age estimates for both pulsars do not practically influence our statistical modeling.

\begin{figure*}[]
\centering
\includegraphics[width=8cm, height= 8cm]{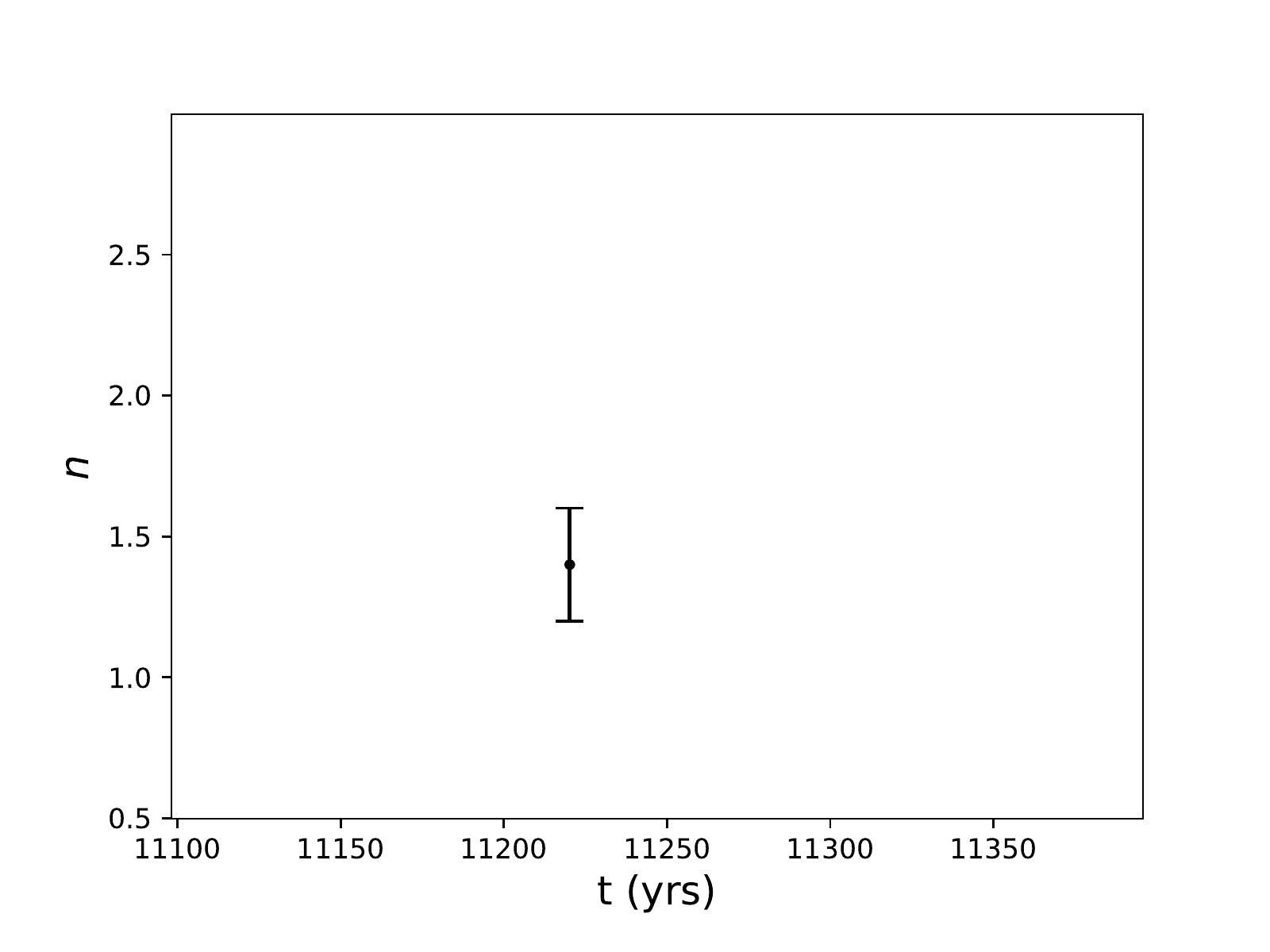}
\includegraphics[width=8cm, height= 8cm]{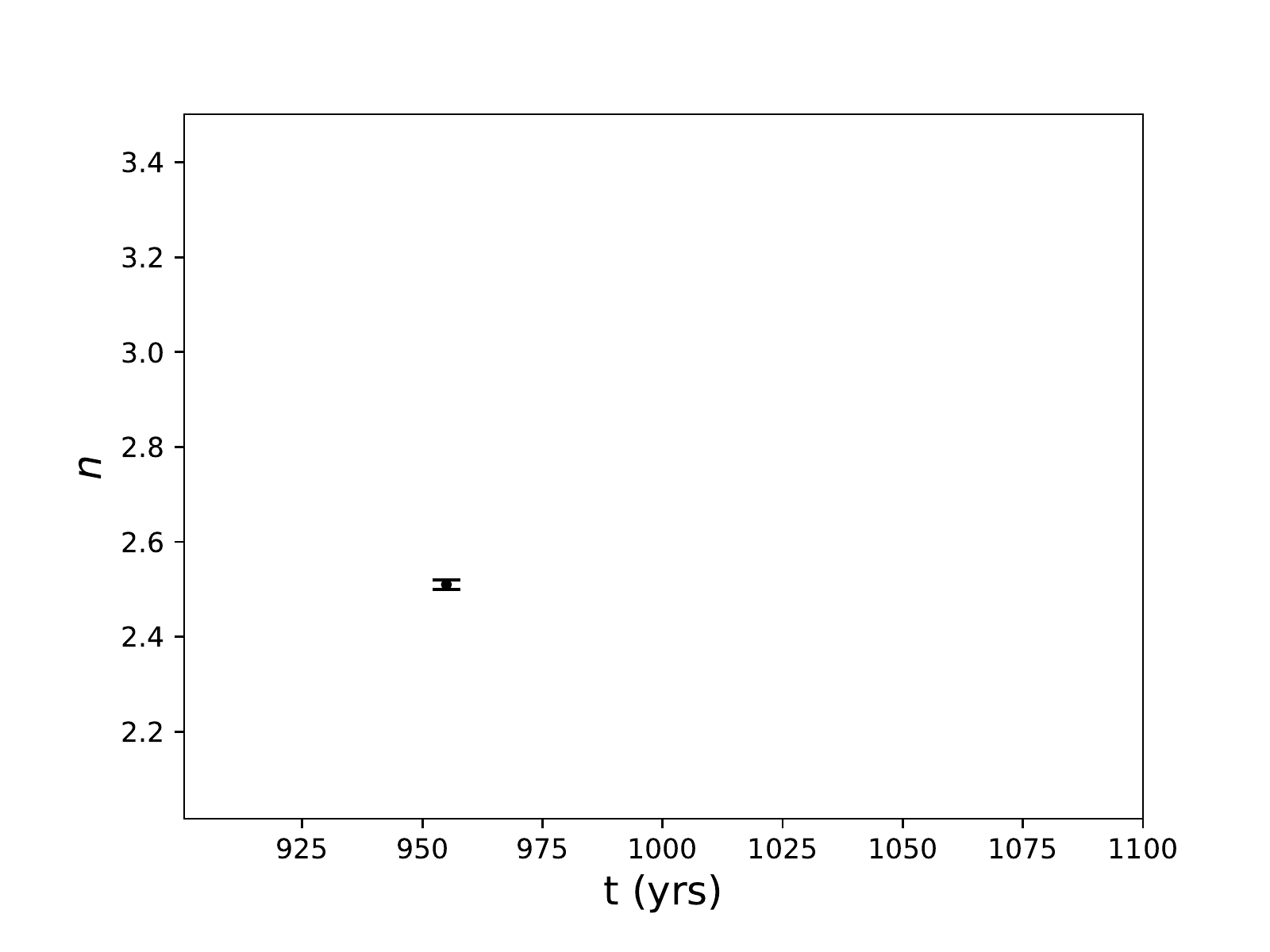}
\caption{Statistical reconstruction at 1$\sigma$ CL of the braking index $n$ as a function of time for Vela and Crab, on the left and right panel, respectively (Case I). The error bar in black represent the $n$ measurements.}
\label{reconstruction_fit_2}       
\end{figure*}

\begin{figure*}[]
\centering
\includegraphics[width=8cm, height= 8cm]{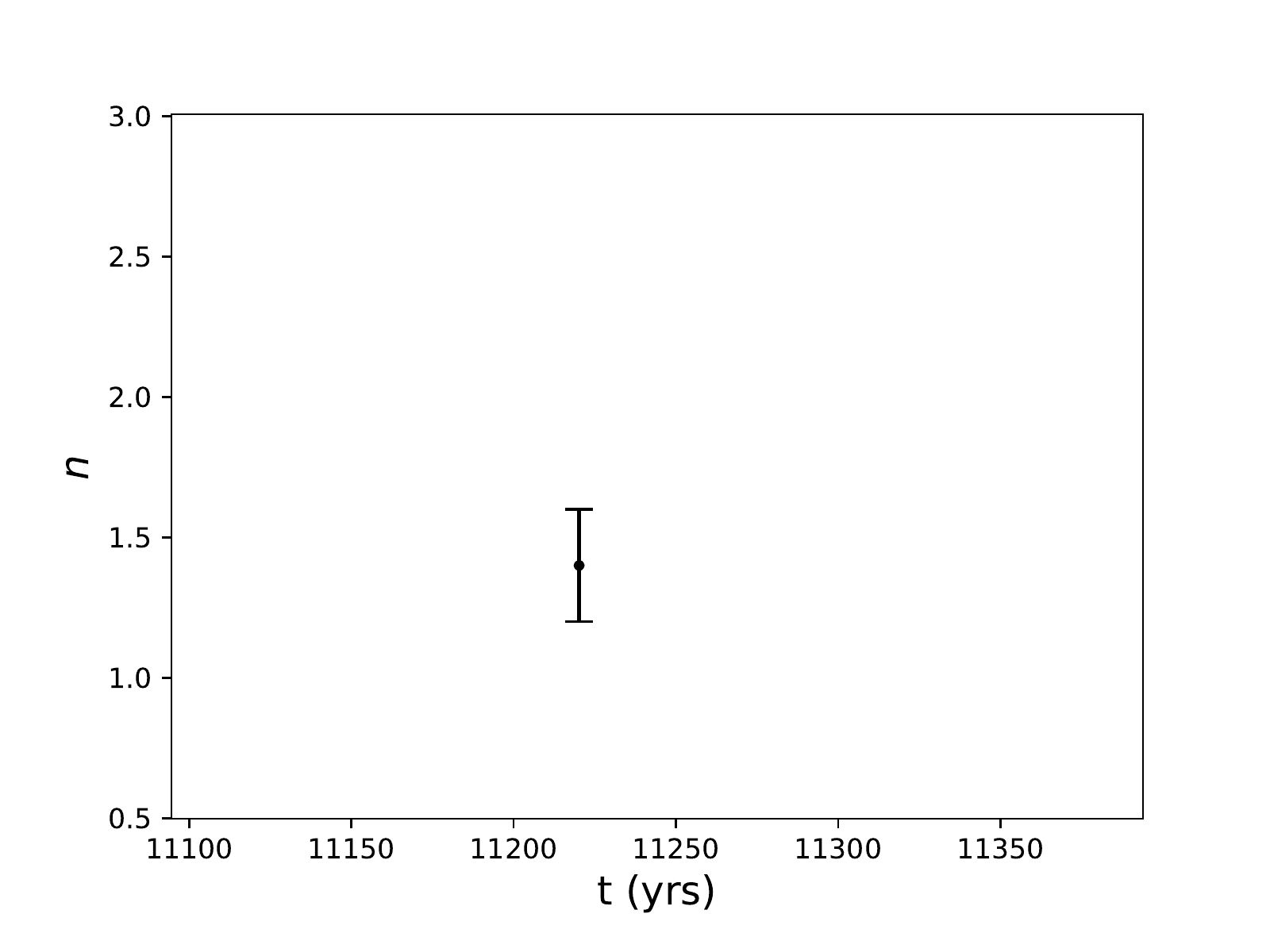}
\includegraphics[width=8cm, height= 8cm]{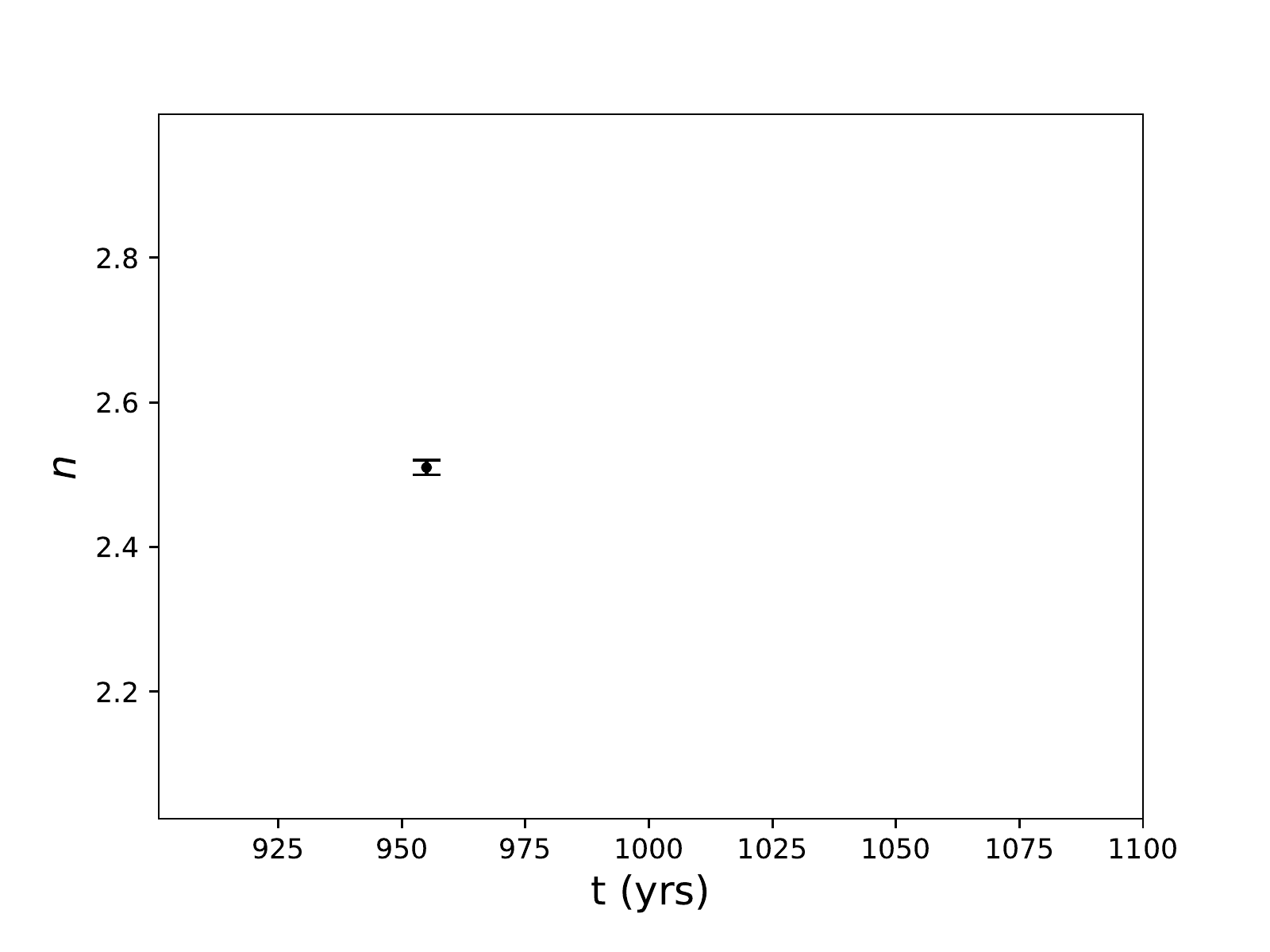}
\caption{Statistical reconstruction at 1$\sigma$ CL of the braking index $n$ as a function of time for Vela and Crab, on the left and right panel, respectively (Case II). The error bar in black represent the $n$ measurements.}
\label{reconstruction_fit}       
\end{figure*}

For the Vela pulsar, we find $\zeta=1280^{+350}_{-630}$ and $\zeta=990^{+320}_{-570}$ at 1$\sigma$ CL from the Case I and II, respectively. For this inference, we find $B = 6.5^{+3.1}_{-4.4} \times 10^{12}$G and $B = 6.8^{+2.5}_{-5.1} \times 10^{12}$G for the the Case I and II, respectively. Now, for the Crab pulsar, we find $\zeta=1002^{+83}_{-76}$ and $\zeta=600^{+160}_{-100}$ at 1$\sigma$ CL from the Case I and II, respectively, with $B = 6.6^{+1.2}_{-1.7} \times 10^{12}$G and $B = 7.3^{+2.1}_{-4.3} \times 10^{12}$G from the case I and II, respectively. Note that the mean value of the $B$ parameter can present statistical fluctuations along the MCMC analysis. But, as expected, these fluctuations are completely compatible with the input value. As previous mentioned the characteristic (inferred) magnetic field from the classical magnetic dipole radiation is subject to some uncertainties. To take into account the magnetic field effects in our results, we have relaxed $B$ using a uniform prior with $B \in [1, 100]$ in units of $10^{12}$G. Nevertheless, it is worth mentioning that up to now, attempts to estimate the magnetic field strength in isolated pulsars through the measurement of cyclotron resonance features, as successfully done for accreting pulsars, have been inconclusive.

Fig. \ref{reconstruction_fit_2} shows the reconstruction at 1$\sigma$ CL of the braking index $n$ as a function of time for Vela and Crab, on the left and right panel, respectively. The reconstruction is done applying standard propagation of error on Eq. (\ref{BI}) from the best fit values obtained in our analysis within the case $\phi \in [45^{\circ}, 70^{\circ}]$. Fig.~\ref{reconstruction_fit} shows the reconstruction for the second case, $\phi \in[70^{\circ}, 90^{\circ}]$. In all of our analysis, we discard the first 10\% steps of the chain as burn-in. We follow the Gelman-Rubin convergence criterion \citep{R_test}, checking that all parameters in our chains had good convergence.

\section{Final remarks}
\label{summary}

There are in the literature several alternatives to the magnetic dipolar brake
to explain the pulsar spindown, among them the pulsar wind model, where a wind of particles coming from the pulsar itself can carry part of its rotational kinetic energy. We have seen that such a spindown mechanism depend critically on three parameters, namely, the dipole magnetic field, the angle between the magnetic and rotation axes, and the density of primary particles of the pulsar's magnetosphere.

Differently from a series of previous articles in this subject, we consider for the first time in the literature a statistical modelling which includes a combination of a dipole magnetic and particle wind brakes. Although in general there is a dependence of all the parameters on the pulsars, we used here, without loss of generality and for the sake of exemplification, only the vacuum gap model for the particle acceleration. We emphasize that this same approach can be applied regardless of the choice of the acceleration model. As a result, we are able to constrain the three relevant parameters of this model, i.e., $B$, $\phi$ and $\zeta$, in  particular for Crab and Vela pulsars.
This study ought to lay the groundwork for future research
on the fundamental parameters of pulsar wind model and also particle acceleration.

\section*{Acknowledgements}
The authors thank the referee for comments which helped to improve the  quality of the manuscript.
J.G.C. is likewise grateful to the support of  CNPq (421265/2018-3 and 305369/2018-0) and FAPESP Project No. 2015/15897-1. J.C.N.A. thanks FAPESP (2013/26258-4) and CNPq (308367/2019-7) for partial financial support. R.C.N. would like to thank the agency FAPESP for financial support under the project No. 2018/18036-5.

\newpage
\bibliographystyle{apj}
\bibliography{ref}

\begin{thebibliography}{40}
\expandafter\ifx\csname natexlab\endcsname\relax\def\natexlab#1{#1}\fi

\bibitem[{{Allen} \& {Horvath}(1997)}]{1997ApJ...488..409A}
{Allen}, M.~P., \& {Horvath}, J.~E. 1997, \apj, 488, 409

\bibitem[{{Bogdanov} {et~al.}(2019){Bogdanov}, {Lamb}, {Mahmoodifar}, {Miller},
  {Morsink}, {Riley}, {Strohmayer}, {Tung}, {Watts}, {Dittmann}, {Chakrabarty},
  {Guillot}, {Arzoumanian}, \& {Gendreau}}]{2019ApJ...887L..26B}
{Bogdanov}, S., {Lamb}, F.~K., {Mahmoodifar}, S., {et~al.} 2019, \apjl, 887,
  L26

\bibitem[{{Chen} \& {Li}(2016)}]{2016MNRAS.455L..87C}
{Chen}, W.-C., \& {Li}, X.-D. 2016, MNRAS, 455, L87

\bibitem[{{Coelho} {et~al.}(2016){Coelho}, {Pereira}, \& {de
  Araujo}}]{2016ApJ...823...97C}
{Coelho}, J.~G., {Pereira}, J.~P., \& {de Araujo}, J. C.~N. 2016, \apj, 823, 97

\bibitem[{{Contopoulos} \& {Spitkovsky}(2006)}]{2006ApJ...643.1139C}
{Contopoulos}, I., \& {Spitkovsky}, A. 2006, \apj, 643, 1139

\bibitem[{{de Araujo} {et~al.}(2016{\natexlab{a}}){de Araujo}, {Coelho}, \&
  {Costa}}]{2016JCAP...07..023D}
{de Araujo}, J. C.~N., {Coelho}, J.~G., \& {Costa}, C.~A. 2016{\natexlab{a}},
  JCAP, 2016, 023

\bibitem[{{de Araujo} {et~al.}(2016{\natexlab{b}}){de Araujo}, {Coelho}, \&
  {Costa}}]{2016ApJ...831...35D}
---. 2016{\natexlab{b}}, \apj, 831, 35

\bibitem[{{de Araujo} {et~al.}(2016{\natexlab{c}}){de Araujo}, {Coelho}, \&
  {Costa}}]{2016EPJC...76..481D}
---. 2016{\natexlab{c}}, European Physical Journal C, 76, 481

\bibitem[{{de Araujo} {et~al.}(2017){de Araujo}, {Coelho}, \&
  {Costa}}]{2017EPJC...77..350D}
---. 2017, European Physical Journal C, 77, 350

\bibitem[{{de Lima} {et~al.}(2020){de Lima}, {Coelho}, {Pereira}, {Rodrigues},
  \& {Rueda}}]{2020ApJ...889..165D}
{de Lima}, R. C.~R., {Coelho}, J.~G., {Pereira}, J.~P., {Rodrigues}, C.~V., \&
  {Rueda}, J.~A. 2020, \apj, 889, 165

\bibitem[{{Du} {et~al.}(2012){Du}, {Qiao}, \& {Wang}}]{2012ApJ...748...84D}
{Du}, Y.~J., {Qiao}, G.~J., \& {Wang}, W. 2012, \apj, 748, 84

\bibitem[{{Dyks} \& {Rudak}(2003)}]{2003ApJ...598.1201D}
{Dyks}, J., \& {Rudak}, B. 2003, \apj, 598, 1201

\bibitem[{{Ek{\c{s}}i} {et~al.}(2016){Ek{\c{s}}i}, {Anda{\c{c}}},
  {{\c{C}}{\i}k{\i}nto{\u{g}}lu}, {G{\"u}gercino{\u{g}}lu}, {Vahdat Motlagh},
  \& {K{\i}z{\i}ltan}}]{2016ApJ...823...34E}
{Ek{\c{s}}i}, K.~Y., {Anda{\c{c}}}, I.~C., {{\c{C}}{\i}k{\i}nto{\u{g}}lu}, S.,
  {et~al.} 2016, \apj, 823, 34

\bibitem[{{Espinoza} {et~al.}(2017){Espinoza}, {Lyne}, \&
  {Stappers}}]{2017MNRAS.466..147E}
{Espinoza}, C.~M., {Lyne}, A.~G., \& {Stappers}, B.~W. 2017, \mnras, 466, 147

\bibitem[{{Foreman-Mackey} {et~al.}(2013){Foreman-Mackey}, {Hogg}, {Lang}, \&
  {Goodman}}]{2013PASP..125..306F}
{Foreman-Mackey}, D., {Hogg}, D.~W., {Lang}, D., \& {Goodman}, J. 2013, \pasp,
  125, 306

\bibitem[{{Gelman} \& {Rubin}(1992)}]{R_test}
{Gelman}, A., \& {Rubin}, D.~B. 1992, Statistical Science, 7, 457

\bibitem[{{Goldreich} \& {Julian}(1969)}]{1969ApJ...157..869G}
{Goldreich}, P., \& {Julian}, W.~H. 1969, {Pulsar Electrodynamics}

\bibitem[{{Gunn} \& {Ostriker}(1969)}]{1969Natur.221..454G}
{Gunn}, J.~E., \& {Ostriker}, J.~P. 1969, \nat, 221, 454

\bibitem[{{Harding} {et~al.}(2008){Harding}, {Stern}, {Dyks}, \&
  {Frackowiak}}]{2008ApJ...680.1378H}
{Harding}, A.~K., {Stern}, J.~V., {Dyks}, J., \& {Frackowiak}, M. 2008, \apj,
  680, 1378

\bibitem[{{Kou} \& {Tong}(2015)}]{2015MNRAS.450.1990K}
{Kou}, F.~F., \& {Tong}, H. 2015, \mnras, 450, 1990

\bibitem[{{Kramer} {et~al.}(2006){Kramer}, {Lyne}, {O'Brien}, {Jordan}, \&
  {Lorimer}}]{2006Sci...312..549K}
{Kramer}, M., {Lyne}, A.~G., {O'Brien}, J.~T., {Jordan}, C.~A., \& {Lorimer},
  D.~R. 2006, Science, 312, 549

\bibitem[{{Landau} \& {Lifshitz}(1975)}]{1975ctf..book.....L}
{Landau}, L.~D., \& {Lifshitz}, E.~M. 1975, {The classical theory of fields}

\bibitem[{{Lander} \& {Jones}(2018)}]{2018MNRAS.481.4169L}
{Lander}, S.~K., \& {Jones}, D.~I. 2018, \mnras, 481, 4169

\bibitem[{Li {et~al.}(2012)Li, Spitkovsky, \& Tchekhovskoy}]{Li_2012}
Li, J., Spitkovsky, A., \& Tchekhovskoy, A. 2012, ApJ, 746, 60

\bibitem[{{Li} {et~al.}(2014){Li}, {Tong}, {Yan}, {Yuan}, {Xu}, \&
  {Wang}}]{2014ApJ...788...16L}
{Li}, L., {Tong}, H., {Yan}, W.~M., {et~al.} 2014, \apj, 788, 16

\bibitem[{{Lyne} {et~al.}(2013){Lyne}, {Graham-Smith}, {Weltevrede}, {Jordan},
  {Stappers}, {Bassa}, \& {Kramer}}]{2013Sci...342..598L}
{Lyne}, A., {Graham-Smith}, F., {Weltevrede}, P., {et~al.} 2013, Science, 342,
  598

\bibitem[{{Lyne} {et~al.}(2015){Lyne}, {Jordan}, {Graham-Smith}, {Espinoza},
  {Stappers}, \& {Weltevrede}}]{2015MNRAS.446..857L}
{Lyne}, A.~G., {Jordan}, C.~A., {Graham-Smith}, F., {et~al.} 2015, \mnras, 446,
  857

\bibitem[{{Lyne} {et~al.}(1993){Lyne}, {Pritchard}, \&
  {Graham-Smith}}]{1993MNRAS.265.1003L}
{Lyne}, A.~G., {Pritchard}, R.~S., \& {Graham-Smith}, F. 1993, MNRAS, 265, 1003

\bibitem[{{Lyne} {et~al.}(1996){Lyne}, {Pritchard}, {Graham-Smith}, \&
  {Camilo}}]{1996Natur.381..497L}
{Lyne}, A.~G., {Pritchard}, R.~S., {Graham-Smith}, F., \& {Camilo}, F. 1996,
  \nat, 381, 497

\bibitem[{{Magalhaes} {et~al.}(2012){Magalhaes}, {Miranda}, \&
  {Frajuca}}]{2012ApJ...755...54M}
{Magalhaes}, N.~S., {Miranda}, T.~A., \& {Frajuca}, C. 2012, \apj, 755, 54

\bibitem[{{Novoselov} {et~al.}(2020){Novoselov}, {Beskin}, {Galishnikova},
  {Rashkovetskyi}, \& {Biryukov}}]{2020MNRAS.494.3899N}
{Novoselov}, E.~M., {Beskin}, V.~S., {Galishnikova}, A.~K., {Rashkovetskyi},
  M.~M., \& {Biryukov}, A.~V. 2020, \mnras, 494, 3899

\bibitem[{{Ostriker} \& {Gunn}(1969)}]{1969Natur.223..813O}
{Ostriker}, J.~P., \& {Gunn}, J.~E. 1969, \nat, 223, 813

\bibitem[{{Padmanabhan}(2001)}]{2001thas.book.....P}
{Padmanabhan}, T. 2001, {Theoretical Astrophysics - Volume 2, Stars and Stellar
  Systems}

\bibitem[{{Riley} {et~al.}(2019){Riley}, {Watts}, {Bogdanov}, {Ray}, {Ludlam},
  {Guillot}, {Arzoumanian}, {Baker}, {Bilous}, {Chakrabarty}, {Gendreau},
  {Harding}, {Ho}, {Lattimer}, {Morsink}, \&
  {Strohmayer}}]{2019ApJ...887L..21R}
{Riley}, T.~E., {Watts}, A.~L., {Bogdanov}, S., {et~al.} 2019, \apjl, 887, L21

\bibitem[{{Ruderman} \& {Sutherland}(1975)}]{1975ApJ...196...51R}
{Ruderman}, M.~A., \& {Sutherland}, P.~G. 1975, \apj, 196, 51

\bibitem[{{Spitkovsky}(2006)}]{2006ApJ...648L..51S}
{Spitkovsky}, A. 2006, \apjl, 648, L51

\bibitem[{{Tong} \& {Kou}(2017)}]{2017ApJ...837..117T}
{Tong}, H., \& {Kou}, F.~F. 2017, \apj, 837, 117

\bibitem[{{Watters} {et~al.}(2009){Watters}, {Romani}, {Weltevrede}, \&
  {Johnston}}]{2009ApJ...695.1289W}
{Watters}, K.~P., {Romani}, R.~W., {Weltevrede}, P., \& {Johnston}, S. 2009,
  \apj, 695, 1289

\bibitem[{{Xu} \& {Qiao}(2001)}]{2001ApJ...561L..85X}
{Xu}, R.~X., \& {Qiao}, G.~J. 2001, \apjl, 561, L85

\bibitem[{{Yue} {et~al.}(2007){Yue}, {Xu}, \& {Zhu}}]{2007AdSpR..40.1491Y}
{Yue}, Y.~L., {Xu}, R.~X., \& {Zhu}, W.~W. 2007, Advances in Space Research,
  40, 1491

\end{thebibliography}

\end{document}